\documentclass[%
 reprint,
 amsmath,amssymb
 aps,
]{revtex4-1}

\usepackage[english]{babel}

\usepackage{graphicx}
\usepackage{dcolumn}
\usepackage{bm}
\usepackage{hyperref}

\newcommand{\eg}{\textit{e.g.}}
\newcommand{\ie}{\textit{i.e.}}
\hypersetup{
    colorlinks=true,
    linkcolor=blue,
    filecolor=magenta,      
    urlcolor=cyan,
}
\begin{document}

\preprint{APS/123-QED}

\title{Real-Time Detection of Unmodeled Gravitational-Wave Transients Using Convolutional Neural Networks}
\author{Vasileios Skliris}
\email{SklirisV@cardiff.ac.uk}
\author{Michael R.~K.~Norman}%
\email{NormanM5@cardiff.ac.uk}
\author{Patrick J.~Sutton}%
\email{SuttonPJ1@cardiff.ac.uk}
\affiliation{%
Gravity Exploration Institute, School of Physics and Astronomy, Cardiff University, Cardiff, United Kingdom CF24 3AA
}%

\date{\today}

\begin{abstract}
Convolutional Neural Networks (CNNs) have demonstrated potential for the real-time analysis of data from gravitational-wave detector networks for the specific case of signals from coalescing compact-object binaries such as black-hole binaries. Unfortunately, CNNs presented to date have required a precise model of the target signal for training. Such CNNs are therefore not applicable to detecting generic gravitational-wave transients from unknown sources, and may be unreliable for anticipated sources such as core-collapse supernovae and long gamma-ray bursts, where unknown physics or computational limitations prevent the development of robust, accurate signal models.
We demonstrate for the first time a CNN analysis pipeline with the ability to detect generic signals -- those without a precise model -- with sensitivity across a wide parameter space. Our CNN has a novel structure that uses not only the network strain data but also the Pearson cross-correlation between detectors to distinguish correlated gravitational-wave signals from uncorrelated noise transients. We demonstrate the efficacy of our CNN using data from the second LIGO-Virgo observing run. We show that it has sensitivity approaching that of the ``gold-standard'' unmodeled transient searches currently used by LIGO-Virgo, at extremely low (order of 1 second) latency and using only a fraction of the computing power required by existing searches, allowing our models the possibility of true real-time detection of gravitational-wave transients associated with gamma-ray bursts, core-collapse supernovae, and other relativistic astrophysical phenomena.

\end{abstract}

\maketitle

\section{\label{sec:intro}Introduction}

Gravitational-wave (GW) astronomy is now an established field of observational science. 
To date, the LIGO~\cite{Aasi_2015} and VIRGO~\cite{Acernese_2015} collaborations have published the details of approximately 90 detection candidates \cite{catO2,2001.01761,2004.08342,2006.12611,PhysRevLett.125.101102,GraceDB,gwtc3} over their first three observing runs.
The detected signals originate from the binary inspiral and merger of two black holes \cite{bbs}, two neutron stars \cite{bns}, or one object of each type \cite{bhns}.

Low-latency detection of candidate signals offers arguably the greatest potential scientific payoff, as the GW observations can trigger followup observations in other channels; \textit{i.e.}, multi-messenger astronomy. 
For example, combined GW and electromagnetic observations of GW170817 - GRB 170817A \cite{GW170817} have yielded novel insights into the origin of heavy elements \cite{Kasen2017}, neutron-star structure \cite{PhysRevLett.121.161101}, GRB astrophysics and host environments \cite{ejecta_kilonova}, and the Hubble constant \cite{hubble}.
Electromagnetic followup of gravitational-wave signals requires very low latency analysis of the GW data - preferably at the second scale to capture the highest energy emissions (e.g., the prompt gamma and x-ray emission of GRBs).
Current low-latency GW analysis techniques rely on hundreds of dedicated CPUs to achieve latencies of tens of seconds to minutes for automated alerts \cite{Abbott_2019}.

Recent work by a number of authors \cite{george_deep_2017, gabbard_matching_2018, bresten_detection_2019, wang_gravitational_2019} has shown that a fundamentally different approach using CNNs has the potential to analyse detector data for GW signals in real time ($\sim$1\,s latency) using a single dedicated GPU. 
However, most CNNs demonstrated to date require a specific signal model for training (\textit{e.g.} an analytic signal model for binary mergers \cite{george_deep_2017, george_deep_2017-1,gabbard_matching_2018, bresten_detection_2019,  wang_gravitational_2019, schmitt_gravitational_2019, gebhard_convolutional_2019, luo_extraction_2019, fan_applying_2019, chatterjee_extraction_2021, deighan_genetic-algorithm-optimized_2021,Huerta2021,HuertaSpining2021,Wei2021},
or catalogs of numerically computed signals for core-collapse supernovae 
\cite{PhysRevD.102.043022,iess2020corecollapse,PhysRevD.98.122002,PhysRevD.103.063011}), and are therefore only capable of detecting signals matching that model.
Many potential sources are governed by physics which is either unknown (\textit{e.g.}~the neutron star equation of state \cite{Steiner:2014pda,BNS_NSradii})
and/or computationally intractable (\textit{e.g.}~the modelling of 
accretion-disk instabilities \cite{disk_instab,van_Putten_2008,PhysRevD.83.044046}); their  transient signals are commonly known as gravitational wave bursts (GWBs).  While the unknown physics governing GWBs makes the study of such signals exciting, it also poses a challenge:
To fully explore the new GW window we need to be able to detect signals from the widest possible variety of sources without relying on precise models for training. 

We address this challenge by proposing a novel CNN architecture that analyses not only the detector strain data directly but also the cross-correlation timeseries between detectors. 
By training the CNN with `featureless' randomised signals, we are able to construct a neural network that detects coherence (amplitude and phase consistency) between detectors rather than specific signal shapes in individual detectors.  
We test our resulting analysis pipeline, which we name \textsc{MLy} (``Emily''), using real data from the LIGO-Virgo network and show that it is capable of detecting a variety of simulated GWB signal morphologies without being specifically trained for them, at sensitivities close to that of standard GWB searches, but at much lower latency and a tiny fraction of the computational cost.

This paper is organised as follows. 
In Section~\ref{sec:background} we give a brief review of applications 
of machine learning in gravitational-wave astronomy.
In section~\ref{sec:cnn} we present the architecture of \textsc{MLy}'s CNNs and describe the analysis procedure and training data.
In Section~\ref{sec:optimizing} we discuss how we optimize the training to maximize performance.
In Section~\ref{sec:results} we present the performance of \textsc{MLy} on both simulated and real LIGO-Virgo data. 
We discuss the implications of these results and next steps in Section~\ref{sec:conclusions}.

\section{\label{sec:background}Machine Learning in Gravitational-Wave Astronomy}

The fields of gravitational-wave astronomy and deep learning have both advanced significantly in recent years. As such, there has been a confluence of research into their combination and a considerable body of work has developed. Artificial neural networks, including CNNs, autoencoders~\cite{shen_denoising_2017}, generative adversarial networks ~\cite{McGinn_2021}, recurrent neural networks ~\cite{PhysRevD.104.064046}, attention-based networks like transformers ~\cite{9956104}, and various other architectures, have been applied to a variety of problems within gravitational-wave astronomy ~\cite{Cuoco_2021}. In this section we summarise a few of these efforts most closely related to the present work; see for example \cite{gebhard_convolutional_2019,Huerta2020,Cuoco:2020ogp} for wider reviews.

Deep-learning studies in gravitational-wave astronomy have most commonly focused on signals from the inspiral, merger, and ringdown of binaries consisting of black holes and/or neutron stars. 
A number of groups have demonstrated such methods for the detection of binary merger signals \cite{george_deep_2017, george_deep_2017-1,gabbard_matching_2018, bresten_detection_2019, Lin2019, wang_gravitational_2019, verma2021employing, PhysRevD.106.084059, schmitt_gravitational_2019, gebhard_convolutional_2019, PhysRevD.103.062004, luo_extraction_2019, fan_applying_2019, chatterjee_extraction_2021, deighan_genetic-algorithm-optimized_2021,Huerta2021,HuertaSpining2021,Wei2021,Ma2024,Jadhav_2023,PhysRevD.107.023021}, as well as for parameter estimation - determining the source properties from a detected signal  \cite{george_deep_2017-1,Chatterjee_2019, shen_deterministic_2019, fan_applying_2019,Gabbard:2019rde,chatterjee_extraction_2021,Chatterjee_2023}.
Deep learning has also been used for fast binary waveform generation \cite{Khan2021}, which could greatly reduce the time required by traditional Bayesian parameter estimation techniques, and for denoising data around signals \cite{shen_denoising_2017,WEI2020135081}. 


In each of these deep-learning studies, the training of the network has relied on the availability of highly accurate models for  
gravitational wave signals from binary mergers \cite{Blanchet_LRR}.
The application of deep learning methods to more general types of gravitational-wave signals has been more limited, with  core-collapse supernovae being the most prominent example.
These studies have made use of either catalogs of gravitational-wave signals from numerical simulations of core collapse or phenomenological waveform models fit to such catalogs.
For example, Chan \textit{et al.}~\cite{PhysRevD.102.043022} trained a CNN using simulated gravitational-wave timeseries with core collapse supernovae signals drawn from a range of published catalogs covering both magnetorotational-driven and neutrino-driven supernovae, and measured the ability to both detect the signal and correctly classify the type.
Iess \textit{et al.}~\cite{iess2020corecollapse} considered the problem of distinguishing true signals from noise fluctuations (``glitches'') that are common in real detectors. They used signals drawn from numerical catalogs combined combined with a simple phenomenological models for two glitch types to train CNNs to distinguish supernova signals from noise glitches.
Lopez~\textit{et al.}~\cite{PhysRevD.103.063011} (building on \cite{PhysRevD.98.122002})
used a phenomenological model mimicking gravitational-wave signals from non-rotating core-collapse supernovae to train 
a complex mini-inception resnet neural network \cite{7780459} to detect supernova signals in time-frequency images of LIGO-Virgo data.

Sasaoka~\textit{et al.} \cite{sasaoka2023deep} \cite{sasaoka2023deepSN} use gradient-weighted feature maps to train CNNs to recognise supernovae spectrograms. They utilised core-collapse supernovae waveforms from a number of catalogues.

Moreno~\textit{et al.} \cite{Moreno_2022} use recurrent autoencoders for anomaly detection. Autoencoders attempt to learn a function to project elements drawn from an input distribution into a dimensionally reduced latent space and then reconstruct the original input element from this reduced latent space. Since the encoder and decoder are trained on a specific distribution, if they are fed an element from outside the distribution, there will be a larger difference between model input and output, indicating an anomaly.

We note that all of these examples, both for binary mergers and for supernovae, rely on having a signal model to train the deep network. 
As a consequence, their applicability is restricted to signals that are similar to the training data.  While not an issue for binary mergers, this may be very important for supernovae where the simulations used for training rely on uncertain physics and numerical approximations and simplifications  \cite{Ott_2009,Abdikamalov2020}. And they are clearly not applicable to the more general problem of detecting gravitational-wave transients from as-yet unknown sources.

Shortly after the release of an early version of this work~\cite{Skliris:2020qax}, Marianer \textit{et al.}~\cite{semisuper} presented a deep-learning algorithm that avoids relying on a signal model by instead using outlier detection. The authors trained a mini-inception resnet network \cite{7780459} on the Gravity Spy data set \cite{Zevin_2017,gravity_spy_2018}, which contains spectrograms of known noise glitches classified into categories. They then applied the CNN to spectrograms of LIGO data and used two methods of outlier detection to identify possible signals. This search was applied to a subset of public LIGO data from the first two observing runs; no signal candidates were found.
To our knowledge this is the only other case to date of a deep-learning method capable of searching for generic gravitational-wave transients.

In this paper we present a deep-learning technique that is capable of detecting generic transient gravitational-wave signals. Our approach differs from previous approaches in a key way: rather than training a CNN to recognise specific signal morphologies in the data streams, we construct CNNs that are designed to recognise \textit{coherence} in amplitude and phase  between two or more data streams. We then train the CNNs using simulated signals and noise glitches that both consist of \textit{random} timeseries with properties drawn from the same distributions. 
Using the same waveform distributions to simulate both the  signals and glitches prevents the CNNs from using the signal morphology to the classify input. 
Instead, the CNNs are forced to learn to measure consistency between detectors.

In the next section we describe the architecture of \textsc{MLy}'s CNNs and the training procedure. We then evaluate \textsc{MLy} by analysing data from the second LIGO-Virgo observing run. We will see that 
our trained pipeline has a detection efficiency approaching that of the standard LIGO-Virgo pipeline for detecting unmodelled gravitational-wave transients \cite{Klimenko:2015ypf}, but with higher speed and lower computational cost.

\section{\label{sec:cnn}A CNN for Unmodelled Burst Detection}

Our goal is to be able to detect sub-second-duration GWBs in data from the three detectors of the LIGO-Virgo network, without prior knowledge of the signal morphology. A significant challenge is to distinguish real signals from the background noise transients, ``glitches'', that are common in these detectors
\cite{o1_allsky,o2_allsky,detcharGW150914}.
Typical GWB detection algorithms
\cite{cWB,olib,x-pipeline,Bayeswave} do this by requiring candidate signals to be seen simultaneously in multiple detectors (simultaneously up to the light travel time between the detectors) and to be correlated between detectors.
We follow this logic in our analysis by using a network architecture for \textsc{MLy} that combines the outputs of two different CNNs: one that detects coincident signals in multiple  detectors (Coincidence Model - Model 1), and a second that detects correlation in phase and amplitude between the detectors (Coherence Model - Model 2). 
Model 1 takes as input the band-passed whitened timeseries data from each detector. 
Model 2 takes the same band-passed whitened timeseries data as well as the Pearson correlation between each pair of band-passed whitened timeseries data [equation (\ref{eq:pearson})]. 
Each model outputs a score on $[0,1]$, where values near 1 indicate a signal and values near 0 indicate noise.
The scores from the two models are multiplied together to give a combined score on $[0,1]$.

In the following subsections, we first review the analysis procedure and the choice of training data. 
We then detail how the architecture for each model is selected.

\subsection{Analysis Procedure}
\label{sec:procedure}

We analyse data from all three of the detectors in the LIGO-Virgo network: LIGO-Hanford (H), LIGO-Livingston (L), and Virgo (V).
In our analysis we use two types of background noise. 
For training we use simulated Gaussian noise that follows the design curves for the LIGO and Virgo detectors \cite{aligo_design}; the motivation for using simulated noise for training is explained in Section~\ref{sec:training}. 
For testing we use real LIGO-Virgo data publicly available from the GW Open Science Center (GWOSC) \cite{open_data_website}. 

The LIGO-Virgo data from GWOSC are sampled at 4096\,Hz. 
We downsample to 1024\,Hz, allowing us to detect signals up to 512\,Hz; this covers the most sensitive frequency range of the detectors and is sufficient for the purpose of demonstrating our CNN. (Since the trained network can process data much faster than real time, we could extend the analysis to higher sample rates. We leave this to future work.)
We chose to focus on signal durations $<$1\ s\, by analysing data in 1\,s segments. This covers many plausible signal models, including for example core collapse supernovae \cite{core_collapse}, perturbed neutron stars and black holes \cite{bhns}, and cosmic string cusps \cite{cosmic_strings}. We could extend to longer durations as well, with a corresponding increase in latency.

The power spectral density (PSD) $S_\alpha(f)$ for each detector $\alpha$ is computed using Welch's method, and used to whiten the corresponding data stream. We use 16 seconds to calculate the PSD, whiten and then keep the central second.
Each data stream is also high-pass filtered at 20\,Hz, giving a search band of [20,512]\,Hz. 
The Pearson correlation of each pair of detectors is given by 
\begin{equation}
\label{eq:pearson}
r_{\alpha\beta}[n] = \dfrac{\sum_{i=1}^N(d_\alpha[i]-\Bar{d}_\alpha)(d_\beta[i+n]-\Bar{d}_\beta)}{\sqrt{\sum_{j=1}^N(d_\alpha[j]-\Bar{d}_\alpha)^2 \sum_{k=1}^N(d_\beta[k]-\Bar{d}_\beta)^2}} \, .
\end{equation}
Here $d_\alpha[i]$ is the band-passed whitened data timeseries for detector $\alpha$, $\bar{d}_\alpha$ is the mean over $N$ samples, and $n$ is an integer time delay between detectors. The correlation is computed for all $n$ corresponding to time delays of up to $\pm$30\,ms with respect to the first detector, which is slightly larger than the maximum possible arrival time difference ($\pm$27.3\,ms) a GW signal can have in the three-detector LIGO-Hanford, LIGO-Livingston, and Virgo network. 

The band-passed whitened data series and the correlation series are then fed into the two models. 
The scores from the two models are multiplied together to give a combined score on $[0,1]$.
In this way a candidate signal needs to score highly for both models; \ie, showing both coincidence in multiple detectors and correlation between the detectors.
In practice we find the scores of both models tend to be strongly peaked around 0 for noise and weak astrophysical signals, and strongly peaked around 1 for strong astrophysical signals.

To estimate the distribution of scores of the background noise, we repeat the analysis many times after time-shifting the data between detectors by an integer number of seconds. Since the time shift is much larger than the largest possible time-of-light delay between detectors, it prevents a real GW signal from appearing in coincidence between multiple detectors. All coincident events in the time-shifted series can therefore be assumed to be uncorrelated and treated as background noise. This is a standard procedure in GW analysis; see \eg~\cite{o1_allsky}. 

To estimate the sensitivity to GWBs, we repeat the analysis after adding simulated  signals to the data. 
In General Relativity, a GW has two polarisations, denoted $h_+(t)$ and $h_\times(t)$. The received signal $h_\alpha(t)$ of a given detector $\alpha$ is the combination
\begin{equation}
    h_\alpha(t) = F_\alpha^+ h_+(t) + F_\alpha^\times h_\times(t)
    \label{eq:response}
\end{equation}
where the antenna response functions $F_\alpha^{+,\times}$ are determined by the position and orientation of the source relative to the detector. We characterise the strength of the received signal by its network signal-to-noise ratio
\begin{equation}  
    \label{eq:rho}
    \rho = \sqrt{ \sum_\alpha ~ 4 \int_0^\infty \frac{|\tilde{h_{\alpha}}(f)|^2}{S_\alpha(f)} \, df } \, 
\end{equation}
Generating simulated signals distributed isotropically over the sky and rescaling to different $\rho$ values allows us to measure the distribution of CNN scores as a function of the signal-to-noise ratio of the signal.

\subsection{Training Data}
\label{sec:training}

The choice of data used to train a CNN is often a critical factor for the CNN's performance.

For the signal population we use white-noise bursts (WNBs) \cite{x-pipeline,o1_allsky}; these are signals where the $h_+$ and $h_\times$ polarisations are independent timeseries of Gaussian noise that is white over a specified frequency range, multiplied by a sigmoid envelope. We select these as our training sample as they are effectively featureless. The bandwidth of each simulated (or ``injected'') signal is selected randomly and uniformly over the range $[40,480]\,$Hz. 
The duration of each injection is selected randomly and uniformly over the range $[0.05,0.9]$\,s. Given their duration their central time is chosen randomly inside this 1-second interval in a way that they aren't cropped.
The injections are distributed uniformly over the sky and projected onto the detectors using equation~(\ref{eq:response}). 
Finally, the signal is rescaled to a desired network signal-to-noise ratio $\rho$ as defined in equation (\ref{eq:rho}). In the rest of this section we will discuss different aspects of our training data and methods, that eventually gave us the current best model.

For the background population we find in practice that the best CNN performance is obtained by training with \textit{simulated} detector noise and glitches, rather than real glitching detector noise. 
The simulated glitches are WNBs with parameters drawn from the same distribution as for GWBs, but independently between detectors (\textit{i.e.}, a different WNB is used for each detector). 
Using simulated background allows us to control the glitch rate in the training set; we will show in Section~\ref{sec:ratios} that we can vary this rate to improve the performance at low false alarm rates. 
Furthermore, using WNBs for the glitches prevents the CNNs from learning to distinguish GWBs from glitches based on the morphology; this is critical since we do not know the true morphology of GWBs.

In the remainder of this section we detail the training process for each model. 
All training data were generated using the \textsc{MLy} package \cite{mly} and its generator function, with elements of the \textsc{PyCBC} \cite{pyCBC} and \textsc{GWpy}\cite{gwpy} package to project the signal onto the detectors and apply time-of-flight differences to the signals arriving at the various detector locations. For training the models we used \textsc{Keras} \cite{keras}.
More details and the codes are available at our repository \cite{scripts} and the mly package ~\cite{mly}.

\subsubsection{Data types}
\label{datatypes}

We have two CNN models to train: the Coincidence Model (model 1) and the Coherence Model (model 2). 
Each model is trained with a dataset chosen to produce the best performance of the model for its assigned task. 

All training samples consist of stationary background noise and optionally an injection into one of more of the detectors. 
The stationary background is Gaussian noise with power spectra that follow the design curves of the LIGO and Virgo detectors \cite{aligo_design}.
For the non-stationary glitch component we use the same WNB waveform family as for the GWB signals but either do the injection into a single randomly chosen detector or we inject into all three detectors but select the WNB parameters independently for each detector.
We use a total of four types of samples:
\begin{itemize}
 \item \textbf{Type 0:} Gaussian LIGO-Virgo noise, with no injections. Labeled as noise.
 \item \textbf{Type 1:} Gaussian LIGO-Virgo noise with coherent WNB injections in all detectors, simulating a real GWB. Labeled as signal.
 \item \textbf{Type 2:} Gaussian LIGO-Virgo noise with different (incoherent) WNB injections in each detector, simulating unrelated glitches or excess noise in each detector. The injections may be simultaneous or offset in time from each other to simulate simultaneous or nearly simultaneous glitches.
 Labeled as noise. 
 \item \textbf{Type 3:} Gaussian LIGO-Virgo noise with a single WNB injection in a randomly chosen detector, simulating a glitch in a single detector. Labeled as noise.
\end{itemize}

The Coincidence Model (model 1) is trained using Type 0 (stationary background), Type 1 (GWBs), and Type 3 (single-detector glitches).
The GWBs are injected with network SNR values in the range [12,30]. The limits of this range are chosen based on the fact that lower SNR values in training increase the false alarm rates dramatically, while training SNRs higher than 30 are not required; we show later that the detection efficiency remains high for SNR values larger than those in the training set. 
Single-detector glitches (Type 3) have SNR values over the range [6,70]. 

The Coherence Model (model 2) is trained using Type 0 (stationary background), Type 1 (GWBs), and Type 2 (incoherent multi-detector glitches).
Experimentation showed that training with GWB network SNRs in the range [10,50] and glitch network SNRs in the range [10,70] gives the best performance. 

The Type 2 incoherent signals represent the extreme cases where glitches are present in the same 1-second interval in all three detectors. 
Real glitches occur independently in different detectors and need not be simultaneous to within the light travel time between detectors.  
To train our model to handle this case we define a quantity called ``disposition'' which is the range of central times of the glitches. 
For example, in the case of three detectors and a disposition $T$ the glitches will be centred at times $T_0-T/2$, $T_0$, and $T_0+T/2$ in the three detectors, with the order selected randomly. The central time $T_0$ is positioned randomly in the 1 second interval so that no signal gets cropped, and the injection durations are also restricted so that no signal gets cropped.
We generate datasets for dispositions distributed uniformly over three different ranges: Type 2a has range of [0.1,0.5] seconds, Type 2b has range [0,0.1] seconds, and Type 2c has zero disposition to challenge the Coherence Model with coincident incoherent signals.

\subsection{Network Architecture}

\subsubsection{Coincidence Model - Model 1}

The first model is a single-input single-output residual neural network whose goal is to identify coincident signals that appear in at least two detectors. This model takes as input the whitened timeseries data from each of the three LIGO-Virgo detectors. The output is a score on $[0,1]$, where high values indicate signal and low values no signal.

Residual neural networks are proven to boost the performance of simpler CNNs by reducing the effect of vanishing gradients; the latter make deep CNNs lose contributions from their first layers and cause their efficiency to saturate. 
We adapt a network from \cite{Ismail_Fawaz_2019} that compares different methods of using machine learning on time-series data and we optimise it using a genetic algorithm (see below).
We find that the resulting residual neural networks outperforms ``ordinary'' deep CNNs in our case.
More specifically we optimise a simple CNN model with a relatively good performance (overall accuracy $>$95\%) and then use it as a “residual block”, where we feed the output of the block to its input. To find the hyper-parameters of the model we use a genetic algorithm \cite{norman} that trains many generations of different randomly initialised models and find which hyper-parameters increase the performance. The two main differences from the original model of \cite{Ismail_Fawaz_2019} are a larger kernel size and the reduction in our residual blocks from three layers to two. We find that three residual blocks are optimal, as in \cite{Ismail_Fawaz_2019}. 
Varying the number of filters has no obvious benefit so we retain the original number to make the model less computationally expensive. 

In training we use early stopping, which stops the training when the loss does not improve for 20 consecutive epochs. 
We find that overall accuracy is further improved 
by using a cyclical learning rate \cite{smith2017cyclical} with the addition that the learning rate is halved when the loss does not decrease for 5 epochs.
Finally, we save the model for every epoch in which the accuracy increases, so that the final model saved is the best one encountered over the training. 

\begin{figure}[h!]
    \centering
    \includegraphics[width=0.45\textwidth]{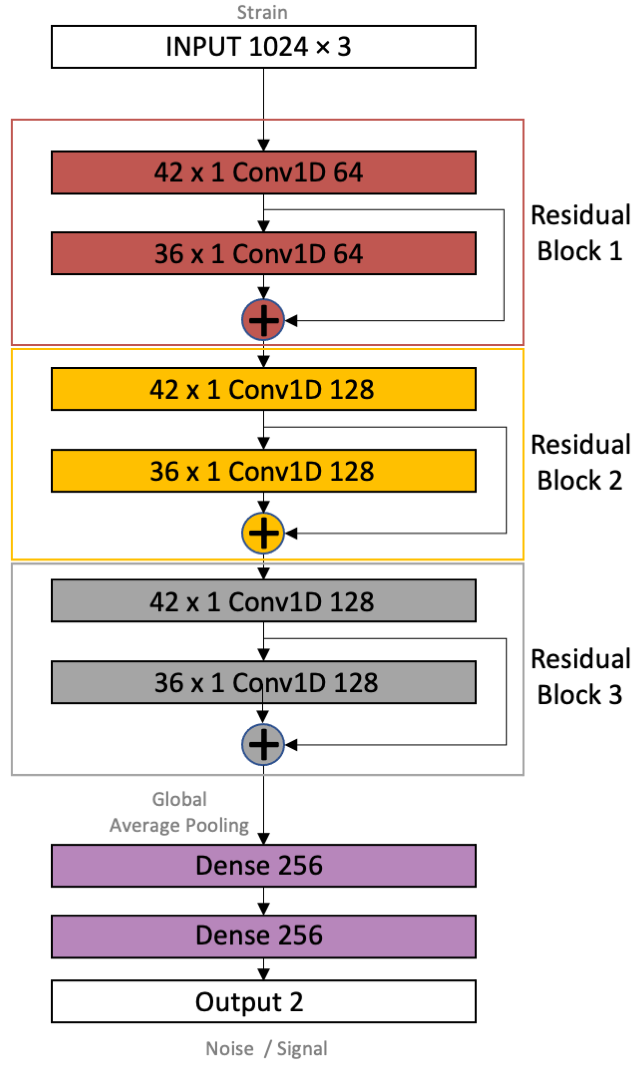}
    \caption{Coincidence model architecture (Model 1). 
    The input consists of three 1024-sample whitened data channels from the H, L, and V detectors. This is passed through three consecutive two-layer residual blocks, flattened, then passed through two final dense layers. $M \times N$ indicates the size of the filters in each convolutional layer and 64/128/256 is the number of filters (convolutional layers) or nodes (dense layers).
    }
    \label{fig:model1}
\end{figure}

The final model is shown in Figure \ref{fig:model1}. It has three residual blocks, the first with 64 filters and the others with 128. At the end of each residual block we add the output of the first layer to the block output. By doing this the gradient can skip the intermediate layer reducing the vanishing effect. Since we use convolution, all layers are zero-padded to maintain the same size with the input and make this addition feasible. At the end of each layer we apply ReLU (rectified linear unit) activation followed by batch-normalisation.
After the last residual block we flatten using global average pooling. We then use two dense (fully connected) layers with batch-normalisation before passing to the output layer, which uses softmax activation. We use binary cross entropy to calculate the loss.

\subsubsection{Coherency Model - Model 2}

The second model has two inputs and one output. The first input is the same whitened timeseries data fed to the first model, while the second input is the Pearson correlation of each pair of detectors, equation (\ref{eq:pearson}).
This model has a binary classification output that is trained to return a measure of coherency among detectors on [0,1]. 

Measurements of correlation between detectors have long been a key ingredient of GWB detection pipelines; see \textit{e.g.}~\cite{cadonati_2004,cadonati_2005,chatterji_etal_2006,x-pipeline,Klimenko:2015ypf}.
We have explored training networks to infer correlation information from the strain data; however, this is not a natural operation for networks constructed from convolutional filters. Since the Pearson correlation is simple to compute and easy to digest by a feature-detecting algorithm, we choose to feed the correlation as a  input to the model.

\begin{figure}[h!]
    \centering
    \includegraphics[width=0.45\textwidth]{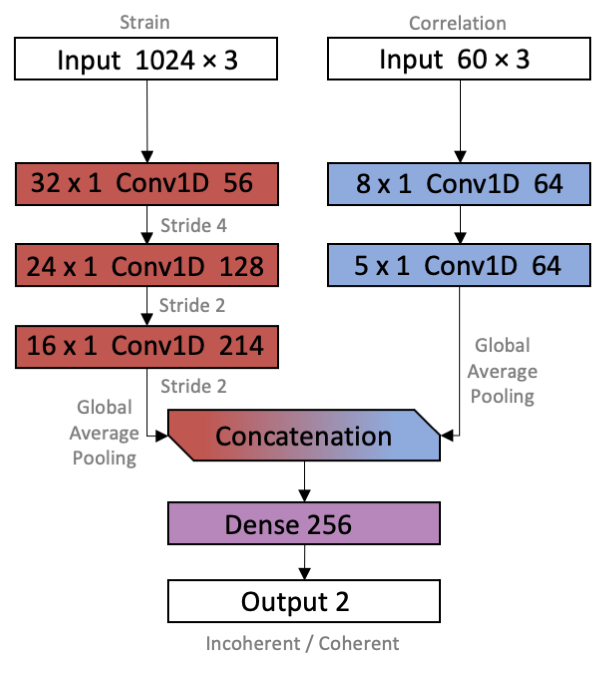}
    \caption{Coherence model architecture (Model 2).
    The model consist of two branches with separate inputs. 
    The input of the left branch consists of the same three 1024-sample whitened data channels as the input to Model 1 (Figure~\ref{fig:model1}). This is passed through three consecutive convolutional layers and flattened.
    The input of the right branch consists of the 60-sample Pearson correlation values [equation (\ref{eq:pearson})] between each pair of whitened data streams. This is passed through two convolutional layers, flattened, then concatenated with the output of the strain branch.
    The combined data is then passed through a dense layer before the output layer. $M \times N$ indicates the size of the filters in each convolutional layer and 64/128/256 is the number of filters (convolutional layers) or nodes (dense layer).}    
    \label{fig:model2}
\end{figure}

The strain and correlation inputs have their own separate branches which are eventually merged as shown in Figure \ref{fig:model2}. 
As for the first model, we use a genetic algorithm to optimise the hyper-parameters. 
Due to the small size of the correlation input, we find that its branch does not need to be as deep as the coincidence model, nor does the strain branch.
We find that the performance is sensitive to the number of filters: increasing or decreasing the number in any of the convolutional layers can prevent the model from training. 
By contrast the kernel size has no significant effect on performance, except for some variation of the stability of training in some cases. The choice of kernel size was made based on how often those numbers appeared in the gene pool of the last generation of successful models in the genetic algorithm.

We use the same early stopping and cyclical learning rate choices as used for model 1.

The final model is shown in Figure \ref{fig:model2}. 
We pass the strain input through three convolutional layers and the correlation data through two convolutional layers in a separate branch. 
(We also explored residual neural networks for the strain branch of the model but but these gave no improvement in performance.)
We flatten the outputs by global average pooling, which increases the performance slightly, and then combine the two branches with concatenation before passing through a dense layer.

\section{Optimizing the Training Procedure}
\label{sec:optimizing}
\subsection{Measuring performance}
\label{sec:measuring}

A standard means to assess the performance of a GWB detection algorithm (see \eg~\cite{o1_allsky}) is to measure the detection efficiency for various signal morphologies as a function of the signal amplitude (\eg, signal-to-noise ratio) at a fixed false alarm rate (FAR). 

We calculate the FAR as a function of score by analysing background data samples that are independent from those used in the training. For the final models we measure the FAR using real LIGO-Virgo data from the second observing run, O2 \cite{o2_allsky}, during times when all three detectors were operating (1 - 25 Aug 2017), with time shifts applied as discussed in Section~\ref{sec:procedure}.

\begin{figure}[t!]
    \includegraphics[width=0.45\textwidth]{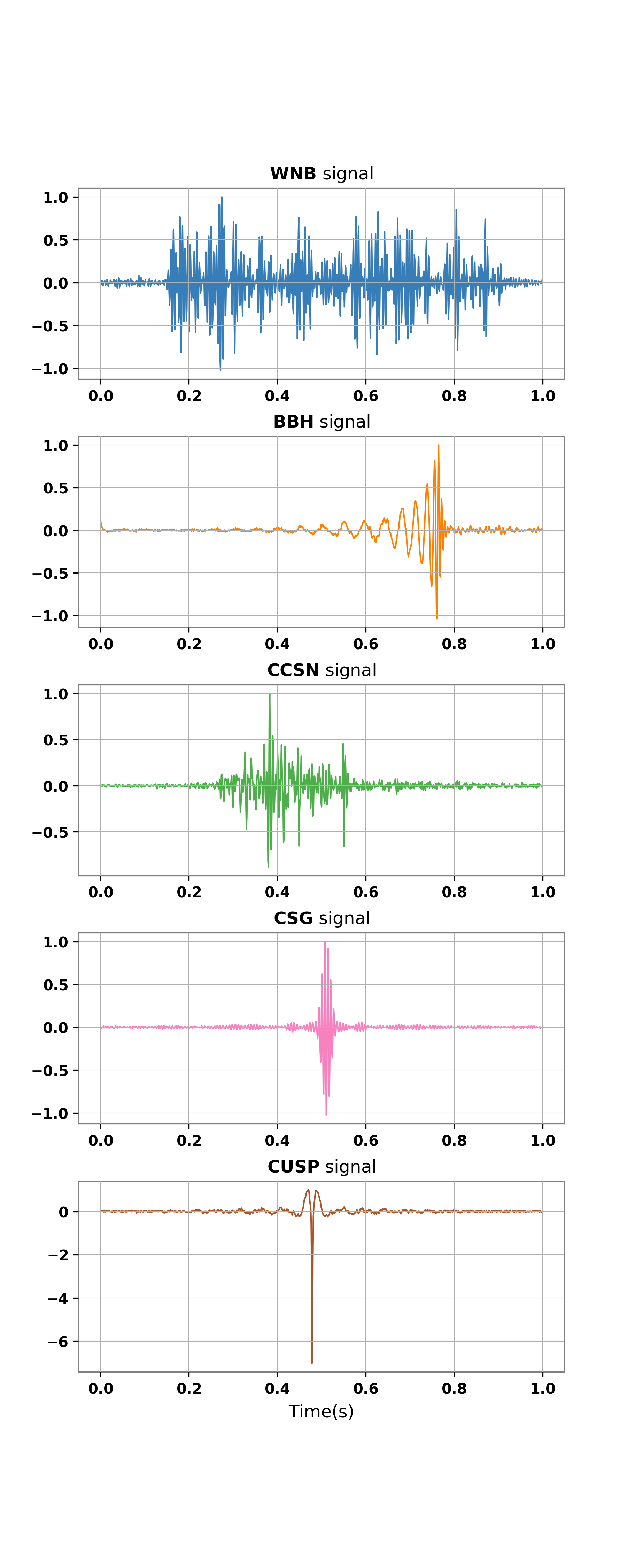}
    \caption{Examples of testing signals: white noise burst (WNB), binary black hole merger (BBH), core-collapse supernova (CCSN), circularly polarised sine-Gaussian (CSG) and cosmic string cusp. 
    All plots show the injection after whitening, but without background noise.}
    \label{fig:waveforms}
\end{figure}

We calculate the detection efficiency for a selection of different possible GWB signals, injected in real LIGO-Virgo data from the second observing run. 
We generate sets of signal injections (different from the ones used for training data) and add them to noise randomly sampled from the O2 observing run during times when all three detectors were operating. We measure our sensitivity to five distinct waveform morphologies:
\begin{description}
\item[WNB] These are the same type of signal as the Type I used for training.
\item[CSG] A circularly polarised sinusoidal signal with Gaussian envelope. These \textit{ad hoc} waveforms are standard for testing GWB analyses~\cite{o1_allsky,o2_allsky}.
\item[CCSN] The N20-2 waveform of \cite{Muller}, from a 3D simulation of a neutrino-driven core-collapse supernova (CCSN) explosion. 
\item[Cusp] The GW emission expected from cosmic string cusps \cite{PhysRevD.64.064008}.
\item[BBH] The GW signal from a black-hole binary merger. 
We use the IMRPhenomD waveform model \cite{IMRPhenomD,PhysRevD.93.044007}.
The black-hole masses and spins are selected randomly and uniformly over the intervals $[10,100]\,M_\odot$ and $[-1,1]$, with the restriction that the maximum signal frequency does not exceed 512\,Hz.
\end{description}
Sample waveforms of each type are shown in Figure~\ref{fig:waveforms}.

We compute the efficiencies at SNR values $\rho=0,1,2,\dots,50$, where the $\rho=0$ case corresponds to pure noise (where we expect approximately zero efficiency).
For each SNR value we generate $100$ injections. 
Each injection is rescaled to the desired network SNR, added to the background timeseries, and processed by the models giving a score. 
The injection is considered to be detected if the combined output score is larger than that of the false alarm rate threshold.

\subsection{Comparing models}
\label{reproducibility}

\begin{figure}[h!]
    \includegraphics[width=0.45\textwidth]{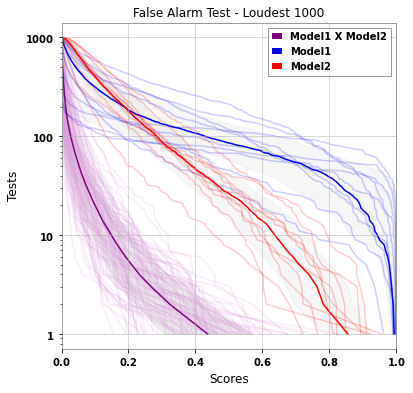}
    \caption{
    False alarm test results for 10 different trainings of model 1 (blue) and model 2 (red) and their 100 combinations (magenta) of scores. The same 1 month of background noise data was used for testing each trained model. 
    We present here the distribution of the 1000 highest background scores for each model. 
    We highlight with a bold line the mean result for each model. The gray area represent one standard deviation for each case.}
    \label{fig:sanitycheck}
\end{figure}

It is crucial for a method to have reproducible results if it is going to be compared with another. Machine learning models typically use random initialisation of their trainable parameters, which can lead to different results for repeated runs of the training. 
As a demonstration of this effect, we train models 1 and 2 ten times each using training data types 0--3, yielding 100 combinations of trained models. 
Figure \ref{fig:sanitycheck} shows the distribution of background scores for each trained version of each model separately and for all 100 model 1 -- model 2 pairs.
We see that while the performance of each model is variable between trainings, the average performance curves are much smoother and more regular.
In the following sections we follow this practice of averaging over repeated trainings of the same model to make more robust comparisons of model performance across different versions of training data.

\subsection{Training sample ratios}
\label{sec:ratios}

In classification problems the training data often contain the same number of examples in each class, giving all classes equal importance.
In our problem we prioritise the reduction of false alarms (false positives) and the ability of the network to recognise noise features. This is motivated by the low false alarm rate threshold adopted by LIGO and Virgo for issuing GWB detection alerts for unmodelled bursts during online running, currently $7.9\times10^{-9}$\,Hz (one per 4 years) \cite{farthr}.
Even with our best model, using equal numbers of signal and noise samples we do not get an acceptable performance. Furthermore we have more than just stationary noise and GWB injections in our data types: we also try to simulate single- and multi-detector glitches, so we need to investigate the optimal proportions of each type. We therefore trained both models with different numbers and ratios of the data types to determine which provide the best performance.

For the Coincidence Model (model 1) we tested all combinations of 3$N$ and 6$N$ instances each of Type 0 and Type 1, and 3$N$, 6$N$, and 12$N$ instances of Type 3, where $N=10^4$, for a total of 12 different combinations. 
For the Coherence Model (model 2) we tested all combinations of 5$N$ and 10$N$ instances each of Type 0, Type 1, Type 2a \& b together, and Type 2c, for a total of 16 different combinations. This gives a total of $12\,\times\,16\,=\,192$ different combinations of training ratios. 

For each combination we repeat the training of each model 7 times and use the mean detection efficiency for comparisons. 
We evaluate the detection efficiency at fixed false alarm rate of 1/day,  using a common set of real background noise data from the second LIGO-Virgo observing run \cite{o2_allsky} and a common set of GWB injections. 
The resulting averaged efficiency curve for a given training ratio combination $i$ and waveform $w$ is characterised by the SNRs at which the efficiency reaches 50\% and 90\%, denoted $\rho^{i,w}_{50\%}$ and $\rho^{i,w}_{90\%}$.
We choose as the best training ratio combination that which minimises the following quantity:
\begin{equation}
\label{eq:ssfe2}
\chi^2_{i} = \sum_{w} \sum_{E=50\%,90\%} \left( \frac{\rho^{i,w}_{E} - \textrm{min}_{j}[\rho^{j,w}_{E}])}{\textrm{min}_{j}[\rho^{j,w}_{E}]} \right)^2
\end{equation}
Minimising this quantity corresponds to achieving a best average performance (lowest $\rho^{i,w}_{50\%}$ and $\rho^{i,w}_{90\%}$) across all waveforms.
We find the best combination to be training model 1 with Types 0, 3, 1 in the amounts 6$N$, 6$N$, 3$N$ and training model 2 with Types 0, 1, 2a\&b, 2c in the amounts 10$N$, 10$N$, 5$N$, 5$N$. This combination gives amplitude sensitivities ($\rho^{i,w}_{50\%}$ and $\rho^{i,w}_{90\%}$ values) 10\%--20\% better than training with equal numbers of each type of data. 
We use these ratios for all training in the rest of this paper.



\subsection{Rescaled Virgo noise level}
\label{sec:virgo}

Up to this point we have trained with artificial Gaussian noise that follows the design noise power spectral density of the advanced LIGO and Virgo detectors \cite{aligo_design}.

\begin{figure}[t!]
    \includegraphics[width=0.45\textwidth]{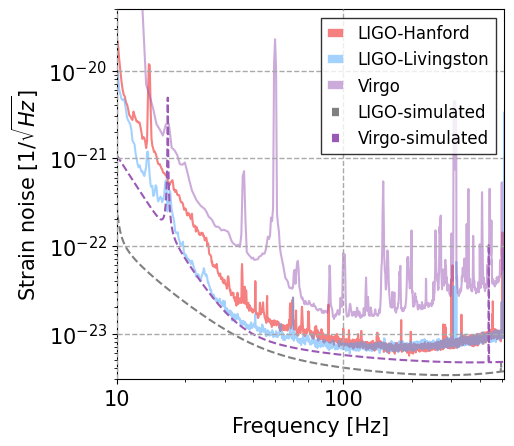}
    \caption{{Noise amplitude spectral densities for the LIGO and Virgo detectors during the second observing run, O2 \cite{o2_allsky}. The dashed lines show the design target sensitivities.}}
    \label{fig:psdo2}
\end{figure}

As shown in Fig.~\ref{fig:psdo2}, when considering the O2 data, the ratio of the true noise level in Virgo to that in the LIGO detectors is higher than the ratio for the design noise spectra used in the training. This means that in the real data we expect the SNR of a signal in Virgo relative to LIGO to be systematically lower than the SNR used in training, potentially resulting in sub-optimal model performance on real data.
To investigate this we retrain the best model from the previous investigation (Section~\ref{sec:ratios}) after rescaling the Virgo PSD by the factors 0.5, 1, 2, 4, 8, 16, or 32. Rescaling by 1 is the same training as before, while the factor of 0.5 is for a sanity check on the method. All other factors ($>1$) lower the SNR of the training signals in Virgo relative to LIGO. For each case we repeat the training 7 times and repeat the false alarm rate and efficiency measurements of the previous section (using new time lags). We then use equation (\ref{eq:ssfe2}) to compare the performance with the different rescalings. 

We find a clear performance improvement when training model 1 with rescaling factors of 2 or 4 and model 2 with rescaling factors of 2 -- 32. The best performance is found using the rescalings (4,32) for models 1 and 2 respectively, giving typical $\rho^{i,w}_{50\%}$ and $\rho^{i,w}_{90\%}$ values almost a factor of 2 lower than training without rescaling. 
We use these rescalings for all training in the rest of this paper.

We note with interest that the optimal scaling factor for training model 2 is larger than that for model 1;. We attribute this to the fact that the orientation of the Virgo detector is very different from the near co-alignment of the two LIGO detectors \footnote{
A simple measure of the overlap of detectors $\alpha$ and $\beta$ is $O_{\alpha\beta} \equiv2 \mathrm{Tr}(\mathbf{D}_\alpha\mathbf{D}_\beta) \in [-1,1]$, where $\mathbf{D}$ is the detector response tensor \cite{PhysRevD.63.042003}. 
One finds $O^{HL}=-0.89$, $O^{LV}=-0.25$, and $O^{VH}=-0.016$.
}; as a consequence, when training with random timeseries, HV and LV correlations tend to be smaller than HL correlations even for equal PSDs.

Rescaling the Virgo noise level upwards lowers the SNR of training signals in Virgo, causing the models to learn to put less emphasis on the Virgo data. 
One might question whether this rescaling means that the Virgo data is not useful.
As a test, we performed a series of injections where each injection was done twice: once normally, with the signal added to all three detector data streams, and again where the same injections were made into Hanford and Livingston only.
We found that while the scores for most injections are largely unchanged, zeroing out the Virgo injection lowers the scores significantly in some cases, particularly for high-SNR signals, while the background distribution is not significantly affected. 
This demonstrates that even with the rescaling of the Virgo noise for training, the models can still extract useful information from the Virgo data stream.

\section{Performance of the Optimally Trained Model}
\label{sec:results}

Following our investigations into optimising the training procedure, 
we now evaluate the performance of \textsc{MLy}'s optimally trained models on real LIGO-Virgo data. 
For these tests we use models trained with the optimal training sample ratios (Section~\ref{sec:ratios}) and with the optimal Virgo noise rescalings  (Section~\ref{sec:virgo}). 
We use new time lags and new injection sets that haven't been used previously. 
Of the 49 combinations of model 1 and model 2 that are trained in this way, we choose the  combination that has the best performance as measured by equation~(\ref{eq:ssfe2}). 
This best combination is then applied to new time lags and injections for the assessments of performance presented in this section.

\begin{figure}[h!]
    \includegraphics[width=0.48\textwidth]{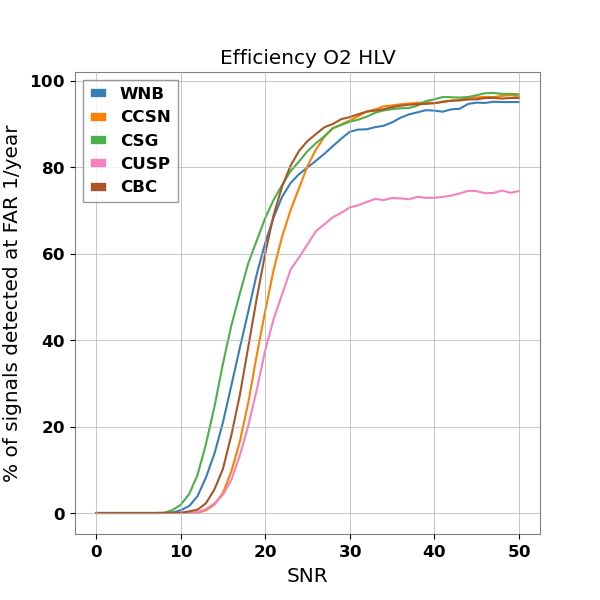}
    \caption{Detection efficiency: the fraction of simulated signals that are detected at a false alarm rate of 1/year versus the network SNR defined by equation~(\ref{eq:rho}). The waveform morphologies are white noise burst (WNB), core-collapse supernova (CCSN), circularly polarised sine-Gaussian (CSG), cosmic string cusp (CUSP), and binary black hole merger (BBH). The background noise is real LIGO-Virgo data from O2.}
    \label{fig:TAR}
\end{figure}

Figure~\ref{fig:TAR} shows the detection efficiency of the optimally trained model for our test waveforms (Section \ref{sec:measuring}) at a FAR of 1/year using real LIGO-Virgo O2 data. We see that \textsc{MLy} is able to detect $>$50\% ($>$90\%) of all signals that have amplitudes $\rho\ge19$ ($\rho\ge27$), with the exception of cusps for which the sensitivity is lower.
The performance for CCSN and BBH signals is very similar to that for WNBs, even though the signal morphologies are quite different (see Figure~\ref{fig:waveforms}). 
The performance for CSGs is even better, which we attribute to the very small time-frequency volume that it occupies making the signal louder. The performance for cusps is poorer; tests indicate that this  occurs because cusps are linearly polarised ($h_\times=0$) while the WNB training injections are unpolarised ($|h_\times|=|h_+|$). 
In the next subsection we will see a similar effect for linearly polarised sine-Gaussian signals. This is an interesting demonstration of where our training could potentially be improved to handle a yet broader range of signals.

\subsection{Comparison with the LIGO-Virgo all-sky search in O2}
\label{sec:comparison}

The LIGO-Virgo collaborations have already searched the O2 data looking for generic short-duration GWB signals \cite{o2_allsky}. We compare the performance of \textsc{MLy} with that of the LIGO-Virgo analysis by repeating the signal injections performed in the LIGO-Virgo search for all signal types reported in \cite{o2_allsky} with maximum frequencies below 500\,Hz; these signal types are listed in Table \ref{tab:table1} and consist of Gaussian pulses, sine-Gaussians, and WNBs (see \cite{o2_allsky} for details).

The procedure to assess efficiencies is the same as that used in previous sections except that, following LIGO-Virgo convention, instead of using SNR, the injected signal strength is characterised by the root-sum-square value $h_\mathrm{rss}$:
\begin{equation}
    \label{eq:hrss}
    h_\mathrm{rss} = \sqrt{\int_{-\infty}^{\infty} \! dt ~ ( |h_+(t)|^2+|h_\times(t)|^2 )}
\end{equation}
Note that $h_\mathrm{rss}$ is independent of the detectors or their noise spectra. Hence for a given $h_\mathrm{rss}$, signals at frequencies of higher detector noise will have a lower SNR. There is therefore no one-to-one match between SNR and $h_\mathrm{rss}$.

\begin{figure}[t!]
    \centering
    \includegraphics[width=0.48\textwidth]{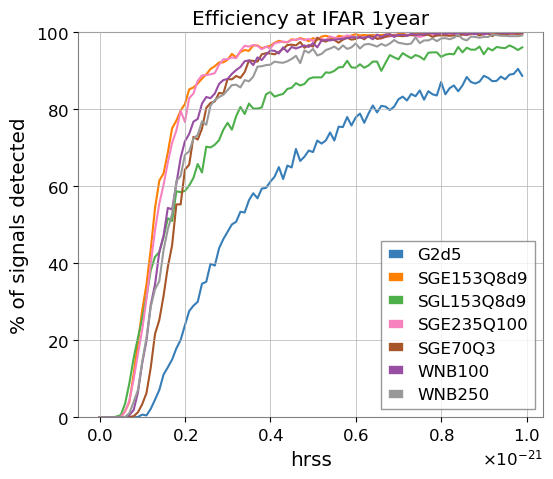}
    \caption{The detection efficiency of the \textsc{MLy} pipeline as a function of the $h_\mathrm{rss}$ amplitude for the burst signal types listed in Table~\ref{tab:table1}. The efficiency is computed at a false alarm rate threshold of 1/year.}
    \label{fig:injo2}
\end{figure}

Figure \ref{fig:injo2} shows the detection efficiency of \textsc{MLy} for the tested waveforms, at a FAR threshold of 1/year. 
We see that most waveforms have similar efficiencies, with only G2D5 and to a lesser extent the SGL153 waveforms having poorer performance. Like the cusp signals in the previous test (Figure~\ref{fig:TAR}), the G2D5 and SGL153 waveforms are the only linearly polarised waveforms in the set. Again we conclude that the lower performance is due to training with exclusively unpolarized waveforms.

Table \ref{tab:table1} reports the $h_\mathrm{rss}$ values at which \textsc{MLy} achieves a detection efficiency of 50\% for each waveform. It also shows the $h_\mathrm{rss}$ values at which the coherent WaveBurst (\textsc{cWB}) \cite{waveburst} pipeline used in the LIGO-Virgo O2 search \cite{o2_allsky} achieves a detection efficiency of 50\% \cite{tiwari}. We see that \textsc{MLy}'s  $h_\mathrm{rss}$ limits are approximately 10\% to 50\% higher than for \textsc{cWB}, corresponding to sensitivity to distances that are 65\% to 90\% as far as those of \textsc{cWB}. 
We consider this to be a very promising first demonstration of the power of machine learning for GWB detection, particularly when one considers the very low computational cost and high speed of the \textsc{MLy} analysis (discussed next).

\newcolumntype{P}[1]{>{\centering\arraybackslash}p{#1}}
\begin{table}[h!]

    \begin{center}

        \renewcommand{\arraystretch}{1.5}
        \begin{tabular}{ P{2.0cm} P{0.7cm} P{3cm} P{1cm} P{1cm} } \hline\hline
            \textbf{Name}  & $\mathbf{+/\times}$ &              \textbf{Parameters} & \textbf{\textsc{cWB}} & \textbf{\textsc{MLy}} \\ 
             & & & \multicolumn{2}{c}{($10^{-22}\,\mathrm{Hz}^{-1/2}$)} \\
            \hline
            \multicolumn{5}{l}{\textit{Gaussian pulses}}\\ \hline
            G2D5 & L & t=2.5\,ms & 2.8 & 3.2 \\ \hline
            \multicolumn{5}{l}{\textit{Sine-Gaussian pulses}}\\ \hline
            SGE70Q3 & E & $f_0$=70\,Hz, Q=3 & 1.5 & 1.9 \\ 
            SGE153Q8D9 & E & $f_0$=153\,Hz, Q=8.9 & 1.3 & 1.4 \\
            SGL153Q8D9 & L & $f_0$=153\,Hz, Q=8.9 & -- & 1.7 \\
            SGE235Q100 & E & $f_0$=235\,Hz, Q=100 & 0.9 & 1.4 \\ \hline
            \multicolumn{5}{l}{\textit{White-Noise Bursts}}\\ \hline
            WNB100 & U & $f_{low}$=100\,Hz, \mbox{$\Delta f$=100\,Hz}, t=0.1\,s & 1.2 & 1.7 \\
            WNB250 & U & $f_{low}$=250\,Hz, \mbox{$\Delta f$=100\,Hz}, t=0.1\,s & 1.4 & 1.7 \\ \hline\hline
        \end{tabular}

    \end{center}

    \caption{Comparison of the detection efficiencies of \textsc{MLy} and \textsc{cWB} at a FAR of 1/year. 
    The first column is the label used for each waveform in Figure \ref{fig:injo2}.
    The second column indicates the signal polarisation: L (linear: $h_\times=0$), E (elliptical: $|h_\times|\le|h_+|$), or U (unpolarised: $|h_\times|=|h_+|$).
    The third column lists the waveform parameters; see \cite{o2_allsky} for definitions.
    The two rightmost columns are the $h_\mathrm{rss}$  values in units of $10^{-22}\mathrm{Hz}^{-1/2}$ at which the \textsc{MLy} and \textsc{cWB} pipelines achieve $50\%$ detection efficiency for each waveform type.   
    A '--' indicates no data available for \textsc{cWB} for that waveform. 
    \label{tab:table1}
}

\end{table}

\subsection{Inference and Training Times}
\label{sec:speed}

We note that the CNN analysis of data is very fast: we find that the average time required to process 1 second of whitened data is 51\,ms on a 3.5\,GHz Xeon E3-1240v5 quad-core CPU with 32\,GB of RAM, or approximately 3.3\,ms on a A100-SXM4-80GB GPU. 
This makes second-scale-latency searches feasible, much faster than the minute-scale latency typical of current LIGO-Virgo low-latency unmodelled burst searches \cite{Abbott_2019}. 

The overall computational cost is also very low compared to traditional search algorithms. 
The dominant computational cost of our analysis is in estimating the FAR. 
Approximately $10^3$ time shifts is enough to estimate accurately the threshold for FAR values of O(1/year) with days to weeks of data;  
three or four A100-SXM4 GPUs will therefore be sufficient to perform the 
full analysis while keeping up with the data in real time. This is in contrast to the several hundred dedicated CPUs typically required by standard algorithms.

Finally, we note that the computational cost of training is modest. A single training of one model on a Tesla V100-SXM2-16GB GPU takes $\sim 1$ hour, and training both models 7 times each takes $\sim 12$ hours.

\section{Conclusions and Future Work}
\label{sec:conclusions}

We have presented a novel CNN-based analysis pipeline, \textsc{MLy}, for the detection of transient gravitational-wave signals. 
Unlike previous CNN-based analyses, \textsc{MLy} is capable of detecting waveforms with morphologies that are not included in the training set while rejecting real detector noise glitches. 
The analysis is shown to be sensitive to a variety of waveform morphologies at signal-to-noise ratios and false alarm rates relevant for issuing rapid alerts to the astronomical community, with very low computing requirements and second-scale latencies possible. 

The \textsc{MLy} pipeline uses a multi-component architecture in which one CNN detects transients that are simultaneous in multiple detectors while a second detects correlation between the detectors to eliminate coincident background glitches. The second CNN takes as input both the whitened detector timeseries data and the Pearson correlation between detectors computed for all physically allowed light travel time delays between detectors, allowing the CNN to detect signal correlation rather than signal shape.
We suggest that using separate models to identify different aspects or properties of the desired signal may be a useful approach generally for GW analysis with machine learning methods.

While our model already has sensitivity approaching that of standard low-latency analyses, we consider this investigation to be a promising first attempt with potential for improvement. 
For example, we could use knowledge of the morphology of common glitch types to reduce the background.  
Our training used simulated glitches with the same morphology as the simulated GWBs to force our models to recognise GWBs by coincidence and correlation between detectors rather than by signal morphology, since the  morphology of real GWBs is not known. However, the morphology of real glitches \textit{is} known. 
Examination of the noise events in the background distribution shows that many of them are due to known glitch types recognised by Gravity Spy \cite{Zevin_2017,gravity_spy_2018,coughlin_classifying_2019,Soni:2021cjy}. Applying a glitch classifier or an auto-encoder to candidate events detected by \textsc{MLy} may allow us to identify false alarms as glitches and thereby veto them in real time. 

While we have demonstrated \textsc{MLy}'s ability to detect GWBs, the characterisation of any detected signals (commonly referred to ``parameter estimation'') is an open problem for unmodelled GWBs, and is necessary for the full exploitation of any detections. 
Existing machine-learning based methods \cite{george_deep_2017-1,Chatterjee_2019, shen_deterministic_2019, fan_applying_2019,Gabbard:2019rde,chatterjee_extraction_2021} (see also \cite{baystar}) are fast but rely on precise signal models for training. 
By contrast the \textsc{BayesWave} algorithm \cite{Bayeswave} is applicable to generic GWBs, estimating the waveform and providing a map of the probability distribution of the source over the sky, but typically requires hours to run for a single event.
Given our goal of using \textsc{MLy} for the low-latency detection of GWBs to allow electromagnetic follow-up observations, a natural next step is to explore how sky-localisation pipelines such as \cite{Chatterjee_2019} can be generalised to the case of unmodelled GWBs. We leave examination of this topic to a future work.

\bigskip
\acknowledgements
We thank Erik Katsavounidis for helpful comments on an earlier draft. 
We thank Shubhanshu Tiwari for providing the sensitivity estimates for \textsc{cWB} listed in Table~\ref{tab:table1}. 
This material is based upon work supported by NSF's LIGO Laboratory which is a major facility fully funded by the National Science Foundation.
The authors are grateful for computational resources provided by the LIGO Laboratory  and supported by National Science Foundation Grants PHY-0757058 and PHY-0823459. 
This work was supported by STFC grants 
ST/N005430/1 and ST/V005618/1.

\bibliography{main}

\end{document}